\documentclass[aps,twocolumn]{revtex4}
\usepackage[dvips]{color}
\usepackage{epsfig}

\usepackage[novolumeabbr]{unitsdef} 
\newunit{\invcm}{\centi\meter\unitsuperscript{-1}}
\newunit{\oersted}{Oe}
\begin{document}

\title{The structure of graphite oxide: Investigation of its surface chemical groups}

\author{D.~W.~Lee,\renewcommand{\thefootnote}{${*}$}\footnote{Corresponding author. E-mail: dongwookleedl324@gmail.com. Phone: +65 6513 8459; Fax: +65 6795 7981.}$^{,}$\renewcommand{\thefootnote}{${\dag}$}\footnote{University of Cambridge}$^{,}$\renewcommand{\thefootnote}{${\ddag}$}\footnote{Nanyang Technological University}
L.~De~Los~Santos~V.,$^{\dag}$ J.~W.~Seo,$^{\dag,\ddag}$ L.~Leon Felix,$^{\dag,\S}$ A.~Bustamante D.,\renewcommand{\thefootnote}{${\S}$}\footnote{Universidad Nacional Mayor de San Marcos} J. M. Cole,$^{\dag,}$\renewcommand{\thefootnote}{${\|}$}\footnote{University of New Brunswick} and C.~H.~W.~Barnes$^{\dag}$}

\affiliation{Cavendish Laboratory, University of Cambridge, JJ Thomson Ave.,Cambridge CB3 0HE, United Kingdom}
\affiliation{Division of Physics and Applied Physics, Nanyang Technological University, 6373616 Singapore}
\affiliation{Laboratorio de Cer\'{a}micos y Nanomateriales, Facultad de Ciencias F\'{\i}sicas, Universidad Nacional Mayor de San Marcos, Ap. Postal 14-0149, Lima, Per\'{u}}
\affiliation{Departments of Chemistry and Physics,
University of New Brunswick, P. O. Box 4400, Fredericton, NB, E3B5A3, Canada}

\begin{abstract}
The structure of graphite oxide (GO) has been systematically studied using various tools such as SEM, TEM, XRD, Fourier transform infrared spectroscopy (FT-IR), X-ray photoemission spectroscopy (XPS), $^{13}$C solid state NMR, and O \emph{K}-edge X-ray absorption near edge structure (XANES). The TEM data reveal that GO consists of amorphous and crystalline phases. The XPS data show that some carbon atoms have $sp^3$ orbitals and others have $sp^2$ orbitals. The ratio of $sp^2$ to $sp^3$ bonded carbon atoms decreases as sample preparation times increase. The $^{13}$C solid-state NMR spectra of GO indicate the existence of -OH and -O- groups for which peaks appear at 60 and 70 ppm, respectively. FT-IR results corroborate these findings. The existence of ketone groups is also implied by FT-IR, which is verified by O K-edge XANES and $^{13}$C solid-state NMR. We propose a new model for GO based on the results; -O-, -OH, and -C=O groups are on the surface.
\end{abstract}

\maketitle

\textbf{1. Introduction}

Graphite is made up of layers of graphene, which is a conductive two-dimensional structure. Since it has a layered structure, it also has a large surface area which renders it suitable for intercalation or adsorption. If it is oxidized in air, CO and CO$_{2}$ are produced. However, if the oxidation is carefully controlled in a laboratory, it becomes insulating graphite oxide (GO). After oxidation, GO still keeps a layered structure with some distortions. Recently, GO has attracted much attention because graphene can be manufactured with GO at low cost\cite{DAN} and GO can be used as a transparent and flexible electronic material after reduction\cite{GOKI} or reaction with NaOH\cite{DOW1}. In addition, it is reported that GO can be used to make metal nanoparticles\cite{Kovtyukhova}.

However, the exact structure of GO still remains elusive despite its various applications. Since it was first discovered in 1859 by Brodie\cite{Brodie}, four different preparation methods have been developed: the Brodie process\cite{Brodie}, the Staudenmaier process\cite{Staudenmaier}, the Hummers-Offeman process\cite{Hummer} and anodic oxidation of graphite electrodes in nitric acid\cite{Hudson-Hunter}. Four main structural models have been proposed for GO: Hofman's model\cite{Hofmann}, Ruess's model\cite{Ruess}, Sholtz \& Boehm's model\cite{Scholz}, and Lerf \& Klinowski's model\cite{Lerf}. Hofmann first proposed that only epoxy (-O-) groups were situated on the surface. Although Hofmann's model explained the existence of epoxy (-O-) groups, it cannot provide the information about other chemical groups including hydrogen. Ruess suggested that the carbon layers were wrinkled and they consisted of trans-linked cyclohexane chairs. He first accounted for the hydrogen content in GO. Ruess's model was revised by Scholz and Boehm of which model says that ketone groups are present in GO and carbon layers were corrugated.

Lerf and Klinowski proposed a new structural model based on solid-state NMR experiments in 1998, and their model has been well received. They assumed that hydroxyl (-OH) and epoxy (-O-) groups exist in GO and carboxyl (-COOH) groups are located at the edges of layers. In spite of the explanation on epoxy and hydroxyl groups, this model leaves some unanswered questions. First, even though the carboxyl groups are present at the edge region in the model, their NMR data do not confirm the existence of carboxyl groups. Second, hydroxyl groups are closely located to each other in the model, causing the electrical instability because of their mutual electrostatic repulsion. For these reasons, Lerf's model needs to be revised, although Lerf $\emph{et al}$\cite{Buchsteiner} modified their model with an explanation of the dynamics of chemical groups in GO. We conducted various experiments to investigate the structure of GO.

\textbf{2. Experimental Methods}

\textbf{A. Preparation} In this work, the GO samples were prepared by the Brodie process\cite{Brodie}. The Brodie process is as follows: 5.0\gram of graphite (99.99+\percent purity, 45\micm, Aldrich) was added into 62.5\milliliter of fuming nitric acid (HNO$_{3}$). After cooling this mixture in an ice bath, 25.0\gram of potassium chlorate (KClO$_{3}$) was slowly added. After the mixture had reached room temperature, it was placed in a water bath, heated slowly to a temperature of 45\celsius and kept at this temperature for 20\hour. Subsequently, the mixture was poured into 125\milliliter of cold distilled water and warmed to 70\celsius, and then the mixture was centrifuged and decanted. The sample was washed three times in this manner and dried overnight at 70\celsius.

\textbf{b. Characterization of the samples} The samples were characterized by powder X-ray diffraction (XRD). XRD patterns were recorded using a Rigaku powder diffractormeter with Cu $K_{\alpha}$ radiation with a step size of 0.02$^{\circ}$. To characterize the shape and structure of the samples, we also used a FEI Philips XL30 sFEG scanning ellectron microscope (SEM), a FEI Philips Tecnai 20 transmission electron microscope (TEM), and a Thermo Nicolet Avatar 360 Fourier Transform infrared spectrometer (FTIR). The elemental compositions of the samples were identified by energy dispersive X-ray (EDX) analysis. X-ray photoemission spectroscopy (XPS) and X-ray absorption near edge structure (XANES) experiments were performed on the BACH beamline at Elettra in Italy. The $^{13}$C solid-state NMR spectra were obtained at 9.4 Tesla using a Bruker AVANCE 400 MHz spectrometer and 4 $mm$ zirconia magic angle spinning (MAS) rotors spun in air at 6 kHz. $^{13}$C MAS spectra with high-power decoupling (HPDEC) were acquired at 100 MHz with 100 kHz 90$^{\circ}$ pulses of 1.25 $\mu$s duration. $^{1}$H-$^{13}$C cross-polarization (CP) MAS spectra were recorded at the different contact times (50 - 5050 $\mu$s) with an increment of 500 $\mu$s.

\begin{figure}[!t]
\centering \leavevmode
\epsfig{file=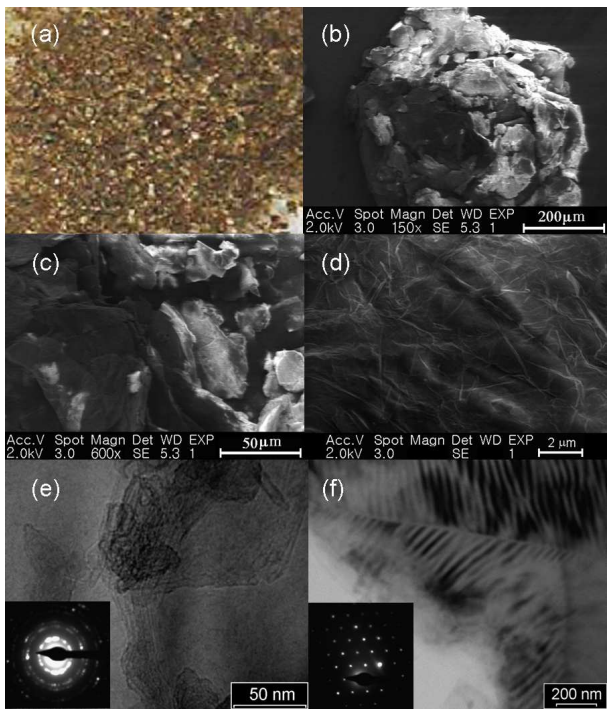, width=8.0cm}
\caption{GO pictures. (a) Optical microscope image; (b)-(d) are SEM images;(e)-(f) are TEM images.}
\end{figure}

\textbf{3. Results and discussion}

Figure 1(a) shows a picture of the sample. It looks dark brown. Figure 1(b)-(d) shows the SEM images of the sample with different magnifications. It looks like coarse powders in Figure 1(b) and (c). It has straight lines and edges in in Figure 1(d). Figure 1e-f show the TEM images of the sample. Figure 1(e) reveals that the sample is made up of layers. The inset is the diffraction pattern of the region. Since the spot-like and circular diffraction patterns appear, there must be two phases: an amorphous phase and a crystalline phase, respectively. Figure 1(f) is another region of the sample. The average width of stripes is 22 $nm$ (9 stripes in 200 $nm$). The stripes appear black and white in turn. The inset shows the crystallinity in the region of Figure 1f. The pattern
is clear and spots are hexagonally arranged, indicating that the region is crystalline. Thus, TEM data confirm that GO is
composed of two phase regions: amorphous and crystalline phases \cite{amorphous-phase}.

EDX analysis were conducted over GO and NaOH-reacted-GO (NaOH-GO)\cite{Epoxy-NaOH}. The results are shown in Figure 2. GO and NaOH-GO show two big peaks. One is from carbon, the other is from oxygen. In addition to these two peaks, NaOH-GO shows a tiny peak from Na. The atomic ratio of C to O for GO is 6:2.2. The ratio of C to O and Na for NaOH-GO is 6:3.2:0.4.

\begin{figure}[!t]
\centering \leavevmode
\epsfig{file=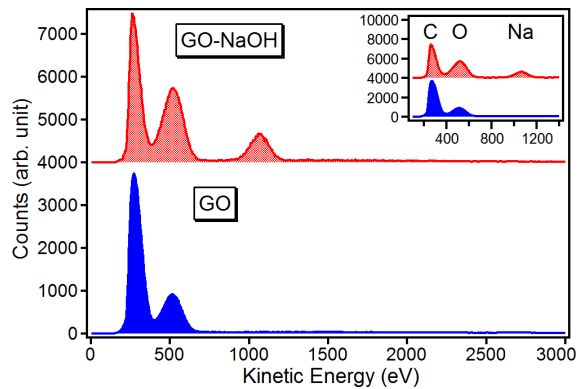, width=7.5cm}
\caption{EDX analysis. (top) NaOH-GO (bottom) GO}
\end{figure}

\begin{figure}[!t]
\centering \leavevmode
\epsfig{file=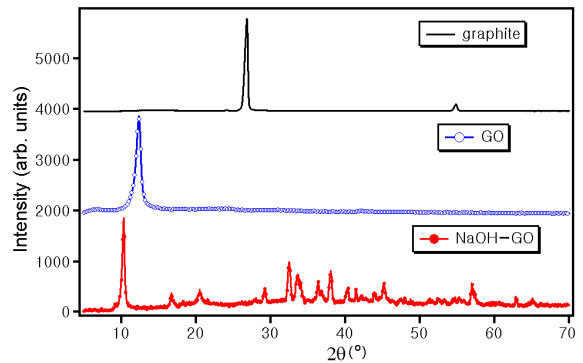, width=7.5cm}
\caption{XRD patterns of graphite, GO, and NaOH-GO.}
\end{figure}

\begin{table}[!t]
\centering
\caption{Principal characteristics obtained by XRD}
\begin{tabular}{c c c c c c} \hline
Sample & 2$\theta$ maximum & FWHM & \emph{D} ($nm$) & \emph{d/2} ({\AA}) & \emph{N} \\ \hline
Graphite & 26.82 & 0.243 & 34.2 & 3.3 & 103 \\
GO & 12.31 & 0.493 & 16.5 & 3.6 & 46 \\
NaOH-GO & 10.29 & 0.257 & 31.6 & 4.3 & 74 \\
\end{tabular}
\end{table}

Figure 3 compares XRD patterns of the samples; graphite, GO, and NaOH-GO\cite{Epoxy-NaOH}. Graphite has the main peak at 26.82$^{\circ}$. After oxidation, the maximum of the main peak moves toward the left, meaning that the interlayer distance increases and the structure is modified. A small amount of diffuse scattering underlies this Bragg signature, implying the onset of a coexistent amorphous phase of GO. After GO has reacted with NaOH, the XRD pattern of NaOH-GO shows many peaks. These additional peaks are due to the structural deformation which occurs by the decomposition of the epoxy groups into hydroxyl and -ONa groups. This decomposition leads to a marked increase in a coexistent amorphous phase of GO, as evidenced by the broad envelope of diffuse scattering that underlies the Bragg peaks. The XRD patterns of the samples were also analyzed with the Scherrer formula\cite{Scherrer}. The full width at half maximum (FWHM), the mean crystallite diameter (\emph{D}), the interlayer distance (\emph{d/2}) and the average number of sheets in the mean crystallite (\emph{N}) are given in Table 1. The interlayer distance increases after the oxidation. It increases further after the reaction with NaOH. This is because more and more chemical groups are formed between the layers after oxidation and the reaction with NaOH. \emph{N} decreases after graphite is oxidized. However, it increases after NaOH decomposes epoxy groups in GO. It might be due to space charges caused by the decomposition of epoxy groups. Oxygen atoms have higher electronegativity than carbon, hydrogen, and sodium atoms (the electronegativity of carbon is 2.55, that of sodium is 0.93, that of hydrogen is 2.20, and that of oxygen is 3.44\cite{Jolly}). Charges transfer from carbon, hydrogen, and sodium atoms to oxygen atoms because charge transfer depends on electronegativity. Sodium and hydrogen atoms bonded with oxygen atoms have positive charges (Na$^{+}$ and H$^{+}$), while oxygen atoms have negative charges(O$^{2-}$). As more hydroxyl groups and -ONa groups are produced between the layers of the sample, more charges are transferred to oxygen atoms and more charges are localized between the layers because GO is an insulator. As a result, \emph{N} might increase due to the electrostatic interaction among the chemical groups.

Figure 4 shows the FT-IR spectra of the samples. There are four main peaks centred at 1050, 1380, 1680 and 3470 $cm^{-1}$. The peak at 1050 $cm^{-1}$ arises from epoxy (-O-) groups. The peak at 1680 $cm^{-1}$ corresponds to the vibrational mode of the ketone (-C=O) groups. The peak observed at 1380 $cm^{-1}$ is assigned to a C-O vibrational mode. The peak at 3470 $cm^{-1}$ denotes C-OH stretching. Peaks below 900 $cm^{-1}$ are not usually interpreted because they represent too complex a structural signature. The peak from epoxy groups in GO becomes weak after reaction of GO with NaOH which can decompose the epoxy (-O-) groups into hydroxyl (-OH) groups and -ONa groups, while the peak from hydroxyl groups grows bigger, revealing that more hydroxyl groups were produced during the decomposition of the epoxy groups. Thus, the FT-IR data verify the existence of epoxy groups in GO and demonstrate that -C=O and -OH groups are present.

\begin{figure}[!t]
\centering \leavevmode
\epsfig{file=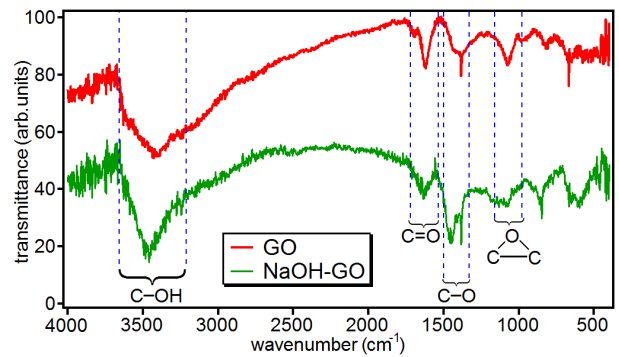, width=8.0cm}
\caption{Characterization of GO and NaOH-GO using FT-IR.}
\end{figure}

\begin{figure}[!b]
\centering \leavevmode
\epsfig{file=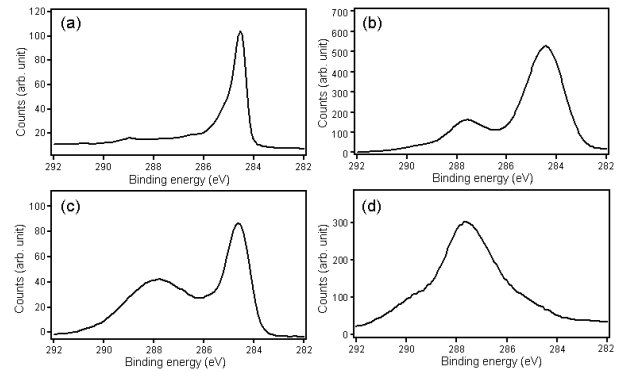, width=8.0cm}
\caption{XPS spectra of the samples; (a) graphite (b) GO (c) NaOH reacted GO for 1 week (d) NaOH reacted GO for 2 weeks.}
\end{figure}

To study what kind of carbon atoms are present in GO, XPS measurements were made on the BACH beamline at Elettra in Italy.
XPS spectra were acquired using a 150 mm VSW hemispherical electron analyzer with a 16-channel detector. The energies of the incident photon were calibrated by measuring the Au 4f photoelectron core level. XPS spectra of the four different samples are exhibited in Figure 5. Graphite has two peaks in Figure 5(a). The peak at 284.5 $eV$ represents C-C bonds with $sp^{2}$ orbitals\cite{Javier}. The small peak at 289 $eV$ is from plasmons\cite{Xie} which is a collective behavior of the delocalized valence electrons of graphite. GO shows two main peaks at 284.5 and 287.6 $eV$ in FIG.~5(b). The first peak corresponds to C-C with $sp^{2}$\cite{Javier}. The other peak originates from C-O in alcohol which is bonded with $sp^{3}$ orbitals\cite{Katzman,Hiroki}. No bending or warping occurs in the regions of carbons bonded with $sp^{2}$ orbitals, whilst bending or warping occurs in the regions of carbons bonded with $sp^{3}$ orbitals. Due to the bending and the coexistence of different orbitals, amorphous phases appear in GO. In reacting with NaOH (Figure 5(c) and Figure 5(d)), more hydroxyl groups are produced. After reacting long enough with NaOH, most epoxy groups ended up to be decomposed into -OH and -ONa groups. The XPS data confirm that epoxy groups in GO decompose to hydroxyl groups and -ONa groups.

To identify in more detail which chemical groups are in GO, $^{13}$C solid-state NMR experiments were conducted. In characterizing solid materials with solid-state NMR, it would be better to compare the high-power decoupling (HPDEC) spectrum of samples with the cross-polarization magic angle spinning (CPMAS) spectrum of them, since we can know which peaks come from carbon atoms near hydrogen atoms by comparing the spectra\cite{Mehring,Taylor,Hansen}. Figure 6(a) shows the $^{13}$C MAS spectrum of GO with high-power decoupling. There are four main peaks at 60, 70, 130 and 195 ppm from tetramethylsilane (TMS). Figure 6(b) illustrates the $^{13}$C CP-MAS spectra of GO recorded using different contact times. The intensity of the peaks increases with contact time. The peaks are the most intense at 2550 $\mu$s, and their intensity decreases afterwards. Figure 6(c) shows the $^{13}$C CPMAS spectrum of GO with the contact time of 2550 $\mu$s, and four main peaks appear at 60, 70, 130 and 195 ppm from TMS. Compared with Figure 6(a), the intensity of the peak at 70 ppm in Figure 6(c) becomes bigger and the peak intensity at 60 ppm in Figure 6(a) is unchanged. Hence, the peak at 60 ppm, which does not cross-polarize, must come from the epoxy group\cite{Lerf}, and the peak at 70 ppm, which is influenced by the nearby hydrogen, must be from the hydroxyl group\cite{Mermoux}. The peaks at 130 ppm are from aliphatic groups\cite{Lerf}. The broad peak at 195 ppm does not cross-polarize. It could be possibly from ketone groups\cite{Masson,Katritzky,Pretsch}. There is no $^{13}$C peak near 175 ppm which comes from carboxyl (-COOH) group\cite{Carboxyl,Bianchi,Knabner}. Since carboxyl groups have hydrogen, its $^{13}$C peak must be cross-polarized if the groups are really present in the sample. Therefore, solid-state NMR spectra indicate that there are epoxy, hydroxyl, and ketone groups in the sample.

\begin{figure}[!t]
\centering \leavevmode
\epsfig{file=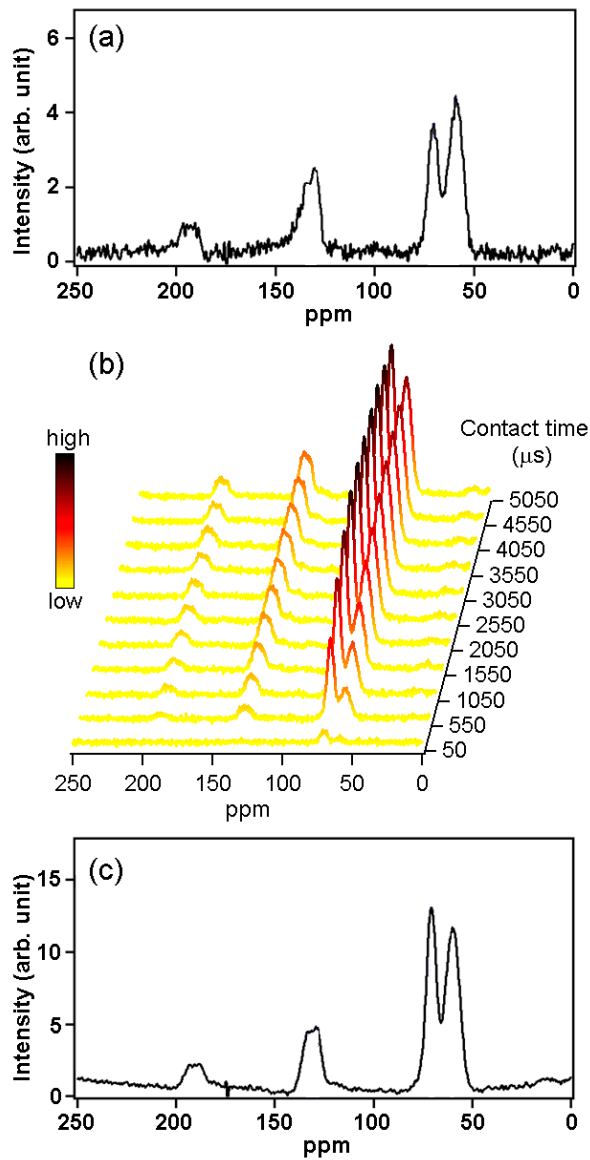, width=8.0cm}
\caption{(a) Solid-state $^{13}$C NMR spectrum of sample GO acquired with high-power decoupling. (b) $^{13}$C CP/MAS NMR spectra of GO acquired using different contact times. (c) $^{13}$C CP/MAS NMR spectrum of sample GO acquired using 2550$\mu$s contact time.}
\end{figure}

It is very important to check whether or not ketone groups are present in GO. Although the peak at 195 ppm in the solid-state NMR spectra does not cross-polarize and the peak position is very similar to that of ketone groups, much care should be taken in interpreting the data and concluding that it is from ketone groups. XANES is one of the powerful techniques to characterize materials. Although it is not easy to examine ketone groups in GO with C $K$-edge XANES since the peak from $\pi_{C=O}$ appears at 285 $eV$ and overlaps with that of $\pi_{C=C}$\cite{Joachim}, O $K$-edge XANES can clearly distinguish ketone groups from other chemical groups. XANES spectra of GO and NaOH-GO were collected in the total electron yield (TEY) mode at room temperature, and they are displayed in Figure 7. Peak A at 532 $eV$ is from -C=O\cite{Reeder,Zemann}. Peak B at 535 $eV$ is from O-H\cite{Hitchcock}. Peak D at 539 $eV$ is assigned to C-O in hydroxyl groups\cite{Naslun}. Since the intensity of Peak C at 537 $eV$ decreases after GO reacts with NaOH, it might originate from C-O in epoxy groups. Therefore, the existence of ketone, epoxy, and hydroxyl groups in GO is confirmed by the XANES spectra.

\begin{figure}[!t]
\centering \leavevmode
\epsfig{file=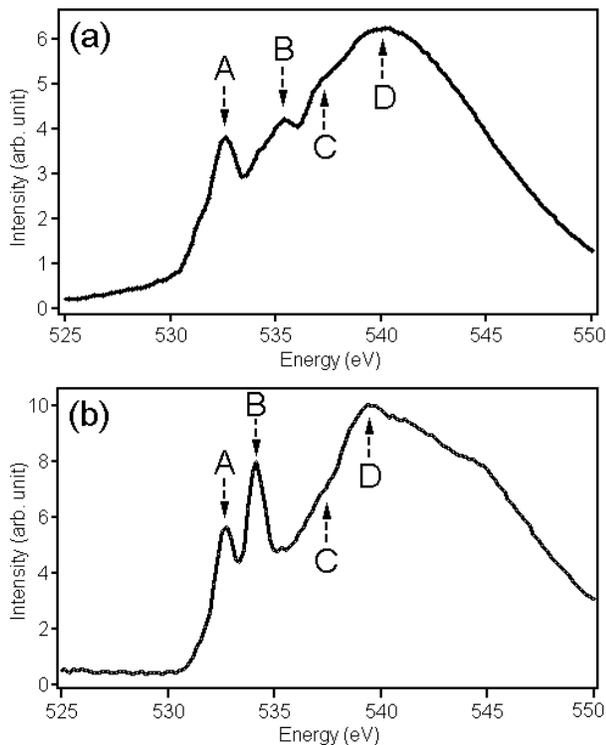, width=8.0cm}
\caption{O \emph{K}-edge XANES spectra of the samples; (a) GO (b) NaOH-GO.}
\end{figure}

\begin{figure}[!t]
\centering \leavevmode
\epsfig{file=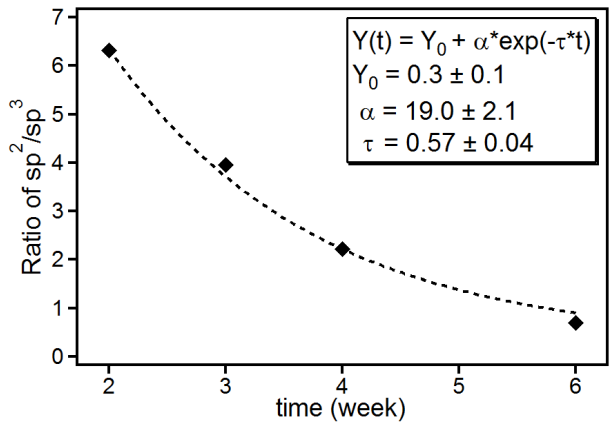, width=8.0cm}
\caption{The ratio of $sp^{2}$- to $sp^{3}$-bonded carbon atoms in the GO samples.}
\end{figure}

Lastly, we investigated whether the ratio of $sp^{2}$ orbitals to carbon atoms with $sp^{3}$-hybridized orbitals in the GO is influenced by sample preparation time. Three more samples were prepared by the Brodie method, and they were characterized by XPS. The peaks of $sp^{2}$ and $sp^{3}$ orbitals were fitted with a Gaussian function, and the ratio is plotted in Figure 8. Figure 8 displays the ratio of carbon atoms with $sp^{2}$ orbitals to carbon atoms with $sp^{3}$-hybridized orbitals in the GO samples against their preparation time. The line in Figure 8 shows the least squares fit of an exponential function. This shows that the ratio asymptotically approaches 0.3 $\pm$ 0.1. This suggests that it is not possible to prepare GO samples with only $sp^{3}$-hybridized orbitals irrespective of the preparation time.

\begin{figure}[!t]
\centering \leavevmode
\epsfig{file=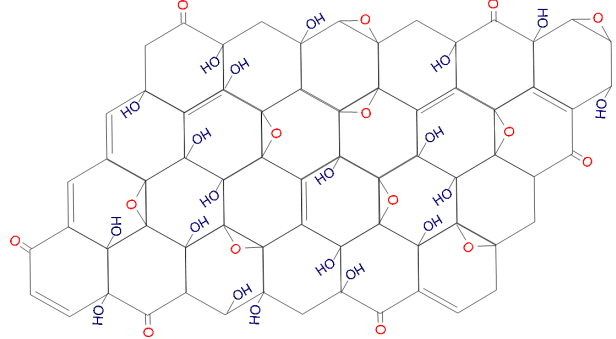, width=8.0cm}
\caption{New structure proposed for GO.}
\end{figure}

Figure 9 proposes the new model for GO in response to all results. As the hydroxyl groups have strong repulsion, they are not located close to each other. Even though they seem to be located randomly in GO, two neighboring hydroxyl groups are located on the opposite sites to decrease the repulsive force between them. Epoxy groups are also separated far from hydroxyl groups in order to be electrically stable. Ketone groups are also present in GO

\textbf{Conclusions}
We investigated the structural properties of graphite oxide (GO) with SEM, TEM, XRD, FT-IR, XPS, $^{13}$C solid-state NMR, and XANES. The TEM data show that GO consists of two regions: amorphous and crystalline. The existence of epoxy, hydroxyl, and ketone groups is corroborated by $^{13}$C solid-state NMR, FT-IR, and XANES spectra. According to XPS data, GO is composed of two kinds of carbon atoms and the epoxy groups are vulnerable to NaOH. We finally propose a new model for GO, and it is illustrated in Figure 9.

\section*{Acknowledgements}
D. W. Lee is grateful to L.~M.~Brown for helpful discussions and is indebted to BACH beamline staff. Luis De Los Santos V. thanks the European Union Programme for Latin America, ALBAN (No E06D101257PE) and Cambridge Overseas Trust for financial support. J. M. Cole acknowledges support from The Royal Society and The UNB Vice-Chancellor's research chair. This paper is written in memory of J. A. C. Bland, who suddenly passed away on 2 December 2007.

\clearpage


\begin{thebibliography}{50}
\bibitem{amorphous-phase} The amorphous phase arises due to the distortion by carbon atoms with $sp^{3}$-hybridized orbitals. Before oxidation, carbon atoms in graphite have $sp^{2}$-hybridzed orbitals and they are on the flat layers. However, some carbon atoms have $sp^{3}$-hybridized orbitals after oxidation and they are not on the layer, causing distortion of layers. As a result, amorphous phase appears where carbon atoms with $sp^{3}$-hybridized orbitals are situated.
\bibitem{Epoxy-NaOH} Epoxy groups are vulnerable to sodium hydroxide (NaOH) and decompose into hydroxyl and -ONa groups. To confirm the existence of epoxy groups in GO, GO was reacted with 0.5M NaOH solution.
\end{thebibliography}
\end{document}